\documentclass[a4paper,11pt]{article}
\pdfoutput=1
\usepackage{jcappub} 
\usepackage{color}
\usepackage{amssymb,amsmath}
\usepackage{graphics,epsfig}
\usepackage{slashed}
\newcommand{\no}{\nonumber\\}
\newcommand{\e}{{\rm e}}

\newcommand{\be}{\begin{equation}}
\newcommand{\ee}{\end{equation}}
\newcommand{\beq}{\begin{equation}}
\newcommand{\eeq}{\end{equation}}
\newcommand{\bea}{\begin{eqnarray}}
\newcommand{\eea}{\end{eqnarray}}
\newcommand{\ba}{\begin{eqnarray}}
\newcommand{\ea}{\end{eqnarray}}

\newcommand{\la}[1]{\label{#1}}

\newcommand{\Tr}[1]{\:{\rm Tr}\,#1}

\title{\boldmath Spectral Regularisation: Induced Gravity
   and the Onset of Inflation} 

\vspace{1cm}
\author[a,b]{Max A. Kurkov,}
\author[c]{Mairi Sakellariadou,}

\affiliation[a]{Dipartimento di Fisica, Universit\`a di Napoli
  Federico II, Naples, Italy} \affiliation[b]{INFN, Sezione di Napoli,
  Italy} \affiliation[c]{Department of Physics, King's College London, University of London, 
  Strand,\\ London WC2R 2LS, United Kingdom}

\emailAdd{max.kurkov@na.infn.it}
\emailAdd{mairi.sakellariadou@kcl.ac.uk}

\abstract{Using spectral regularisation, we compute the Weyl anomaly and express the anomaly generating functional of the quantum effective action through a collective scalar degree of freedom of all quantum
vacuum fluctuations.  Such a formulation allows us to describe induced gravity on an equal footing with the anomaly-induced effective action, in a 
self-consistent way. We then show that requiring stability of the cosmological constant under loop quantum corrections, Sakharov's induced gravity
and Starobinsky's anomaly-induced inflation are either both present or both absent, depending on the particle content of the theory.
}

\begin{document}
\begin{flushleft}
KCL-PH-TH/2013-38
\end{flushleft} 

\maketitle
\flushbottom
\section{Introduction}\label{sec:intro}
It is widely believed that the early stages of our universe were
characterised by a phase of an accelerated expansion, dubbed
inflation~\cite{Starobinsky:1980te,Guth:1980zm,Linde:1981mu}, during which the length scales increased by approximately $e^{75}$ times within a relatively short time of less than about $10^6 t_{\rm Pl}$ (with $t_{\rm Pl}$ denoting the Planck time).  To realise such an inflationary era, one usually requires particular initial conditions at the end of the quantum gravity era~\cite{Calzetta:1992gv,Calzetta:1992bp,Germani:2007rt} and 
the presence of a (scalar) field, called the inflaton, whose potential satisfies certain properties. It would certainly be more natural to
have one of the scalar fields of the theory playing the r\^ole of the
inflaton and the scenario of Higgs-driven
inflation~\cite{Bezrukov:2007ep} even though appealing, it faces several
difficulties~\cite{Burgess:2009ea,Barbon:2009ya,Buck:2010sv,Atkins:2010yg}. Going beyond the minimal Standard Model, there are also difficulties in finding a natural candidate for the inflaton field, as it has been recently shown for the case of hybrid inflation\footnote{Hubrid inflation seems also to require fine tuning of its free parameters~\cite{Rocher:2004et,Rocher:2004my,Rocher:2006nh,Battye:2010hg}    since such a model leads generically to cosmic string formation~\cite{Jeannerot:2003qv}, whose contribution to the Cosmic Microwave Background (CMB) temperature anisotropies is heavily constrained from the data~\cite{Ade:2013xla}.}
within supersymmetric SO(10)~\cite{Cacciapaglia:2013tga}. 
Hence usually one assumes the presence of an additional --- probably within the hidden sector --- scalar field with certain appropriate properties --- flatness of its potential and fine-tuned couplings to accommodate the CMB data --- so that it can lead to a successful inflationary era. This approach, even though is by construction successful, it is nevertheless not fully satisfactory. Hence, alternative approaches have been proposed and studied in the literature.  

One can, for instance, add an $R^2$-term in the gravitational action, without changing the particle content of the theory~\cite{Kofman,Gottlober:1990um}. Such a term could appear
in the context of quantum gravity corrections as one approaches the Planck scale. It is clearly appealing to study the realisation of an inflationary era through the effects of Quantum Field Theory (QFT) in curved space-time, while keeping the required phenomenological input to its minimum; this is the philosophy we will adopt in this study.

During the very early stages of our universe, matter can be described by a set of massless fields with negligible interactions.  Such fields, studied in the context of QFT in curved space-time, may lead
to an inflationary era. More precisely, trace (conformal) anomaly,
resulting from the renormalisation of the conformal part of the vacuum
action, becomes the dominant quantum effect and can drive an
inflationary era in the absence of an inflaton field. Such a proposal was first introduced by Starobinsky~\cite{Starobinsky:1980te}, then studied by
Vilenkin~\cite{Vilenkin:1985md} and more recently it has been further investigated by various authors  (see for instance Ref.~\cite{Shapiro:2003gm,Pelinson:2002ef} and references therein, and Refs.~\cite{Pelinson:2010yr,Fabris:2011qq}). 

Even though the Einstein-Hilbert action can be also seen as an induced quantum gravity effect~\cite{Sakharov:1967pk,Visser:2002ew}, one cannot --- to our knowledge --- find in the literature a consistent mathematical scheme allowing to describe simultaneously induced quantum gravity and
anomaly-induced effective action. Standard computation of trace
anomaly in curved space-time usually relies on  $\zeta$-functional regularisation~\cite{jeta-functional}, that does not exploit the ultraviolet cutoff scale, thereby missing the effect of Sakharov's induced gravity. In contrast to that, the frequently used Schwinger's proper time regularisation~\cite{Schwinger}, that gives immediately the Einstein-Hilbert action as a quantum effect, is not suitable to describe the 
Weyl anomaly, since it leads to a local Weyl noninvariant expression, while anomaly generating functional is however known to be nonlocal~\cite{Riegert:1984kt}.
 
Our aim here is to investigate, in a  self-consistent way, whether a model whose only ingredient is QFT can generate a successful inflationary era and moreover imply an induced Einstein-Hilbert action.  We will thus revise the onset of inflation driven by the trace anomaly of the quantum effective action, in the absence of both, a ``bare'' action for gravity and an inflaton field, in the sense of a scalar field incorporated in the model. In other words, we will investigate whether a cosmological arrow of time can 
result from a purely quantum effect. We  will apply spectral regularisation
in a classical conformal invariant theory. We will neglect any effects of masses at high energies, supposing that they are all much smaller than the corresponding value of the Hubble parameter.  We will show that  spectral regularisation allows us to derive the Weyl anomaly generating functional describing both the induced Einstein-Hilbert action and the standard Weyl anomaly. However, spectral regularisation is highly nontrivial.  The price we have to pay for capturing the two effects simultaneously, is the introduction of an auxilary field, that can be considered as a collective degree of freedom of vacuum fluctuations of all fields, dual to conformal anomaly, that we call ``collective dilaton". 

Spectral regularisation was first introduced in a context of chiral, and later also, scale anomalies in Quantum Chromo Dynamics (QCD) in flat space-time~\cite{Andrianov:1983fg, Andrianov:1987jh}.  It was later applied in the case of fermions moving in a fixed bosonic background in the context of induced gravity and the collective dilaton Lagrangian was computed up to linear order in curvature~\cite{Vassilevich:1987yn}. More recently, it was extended up to quadratic order in curvature in the context of the bosonic spectral action~\cite{Andrianov:2011bc,Kurkov:2012dn}. In the present study, we generalise the spectral regularisation for all quantised fields, in order to study in a systematic way the influence of quantum vacuum fluctuations on the gravitational dynamics.

This paper is organised as follows: In Section~\ref{sec:anomaly} we
derive a mathematical description of anomaly using spectral
regularisation.  In Section~\ref{sec:indgrav} we show how the induced
gravitational Einstein-Hilbert action appears in our
formalism. Section~\ref{sec:infl} is devoted to trace anomaly induced
inflation. We will show that, requiring stability of the cosmological
constant under loop corrections, the condition of Sakharov's induced
gravity becomes equivalent to the condition for the existence of a
stable inflationary solution.  We round up with our conclusions in
Section~\ref{sec:conclusions}. Finally, some technical aspects are presented in the Appendix. 

\section{Spectral Regularisation and Collective Dilaton Lagrangian 
\la{sec:anomaly}}

We will derive a mathematical description of anomaly using spectral
regularisation.  We will first compute the anomaly and then present the anomaly generating functional. The latter is achieved through the introduction of an
auxilary field, that can be considered as a collective degree of
freedom of vacuum fluctuations of all fields, dual to conformal
anomaly.

\subsection{Spectral Regularisation: A Brief Overview}

Our main aim is to compute the influence of vacuum fluctuations of
quantised fields on the dynamics of the metric tensor in the context
of QFT with an ultraviolet cutoff.

Since in asymptotically free QFT, the interactions --- non-abelian
interactions, Yukawa interactions and Higgs self-interactions --- can
be considered as perturbative, the effect we are interested in is, at
leading order, given by one-loop vacuum energy of free fields.
However, even this simple approximation may lead, in curved
space-time, to non-trivial effects like Sakharov's induced
gravity~\cite{Sakharov:1967pk} and Starobinsky's anomaly-induced
inflation~\cite{Starobinsky:1980te}.

Let us consider a theory of free quantised fields of various
spins moving in a gravitational background. The classical action reads
\be
\label{clas-act}
 S_{\mathbf{cl}} =\int {\rm d}^4 x \sqrt{g}\left(
\sum_{j=1}^{N_{\mathcal H}} ~{\mathcal H}_j D_{\mathcal H} {\mathcal
  H}_j + \sum_{j=1}^{N_{\rm F}}~\bar\psi_j \slashed{D} \psi_j +
\frac{1}{4} \sum_{j=1}^{N_{\rm V}} ~ F_{\mu\nu}{}_j
F^{\mu\nu}{}_j\right)~,
\ee
where
\ba
\slashed{D} &=&i
e_k^{\mu}\gamma^k\left(\partial_{\mu} -
\frac{i}{2}\omega_{\mu}^{mn}\sigma_{mn} \right)~,\\ 
D_{\mathcal H}&=& -\nabla^2 -\frac{R}{6}~, \ea
with $F_{\mu\nu}$ the field strength, $R$ the scalar curvature,
$\omega_\mu^{m n}$ the spin connection and $\sigma_{m n}$ the
generators of the representation of the Lorentz group.  Note that
$N_{\rm F}, N_{\rm V}, N_{\mathcal H}$ stand for the number of Dirac
four component fermions, gauge vector bosons and real Higgs-like
scalars, respectively. The classical action Eq.~(\ref{clas-act}) is conformally
invariant. This setup may be considered as a good description of the
Standard Model (or its generalisations) when all masses are much smaller than the Planck mass
and the scalar fields are conformally coupled to gravity.

In order to quantise the theory we follow Faddeev-Popov gauge fixing
procedure. In Feynman-t'Hooft gauge (a type of an $R_\xi$ gauge, as a
generalisation of the Lorentz gauge, with $\xi=1$), the action reads
\be S_{\mathbf{cl,gf}} =\int d^4 x
\sqrt{g}\left[\sum_{j=1}^{N_{\mathcal H}}~{\mathcal H}_j D_{\mathcal
  H} {\mathcal H}_j + \sum_{j=1}^{N_{\rm F}}~\bar\psi_j \slashed{D}
\psi_j + \sum_{j=1}^{N_{\rm V}} \left(\frac{1}{2}A^{\mu}{}_j
\left(D_{\mathbf{vec}}\right)_{\mu}^{\nu}A_{\nu}{}_j + \bar c_j
D_{\mathbf{gh} } c_j\right)\right]~, \ee
where
\ba
\label{D_gh}
D_{\mathbf{gh}}  & \equiv& -\nabla^2~,\\
 \left(D_{\mathbf{vec}}\right)^{\nu}_{~\mu} &\equiv& -
\delta_{\mu}^{\nu}\nabla^2 - R^{\nu}_{\mu}~. \ea
The object we are interested in, is a quantum partition function that
(up to irrelevant constant) is given by
\ba Z &\equiv& \int[{\rm d}\bar \psi][{\rm d}\psi][{\rm d} {\mathcal
    H}][{\rm d} A][{\rm d}\bar c][{\rm d} c]
e^{-S_{\mathbf{cl}}[\bar\psi,\psi,{\mathcal H},A,\bar c,
    c,g_{\mu\nu}]} \nonumber\\ &=&Z_{{F}}^{N_{\rm F}}\cdot
Z_{{\mathcal H}}^{N_{\mathcal H}}\cdot Z_{{\mathbf{vec}}}^{N_{\rm
    V}}\cdot Z_{{\mathbf{gh}}}^{N_{\rm V}}~,
\ea
and is formally equal to:
\be
Z=\frac{\left({\det\left(\slashed{D}^2\right)}\right)^{\frac{N_{\rm
        F}}{2}}} {\left({\det\left(D_{\mathcal
      H}\right)}\right)^{\frac{N_{\mathcal H}}{2}}}
\frac{\left({\det\left(D_{\mathbf{gh}}\right)}\right)^{N_{\rm V}}}
     {\left({\det\left(D_{\mathbf{vec}}\right)}\right)^{\frac{N_{\rm
             V}}{2}}}~. \la{Ztot} \ee
Note that in a theory with $N^{\mathbf w}_{\rm F}$ two-component Weyl
fermions one should replace $N_{\rm F}$ by $N^{\mathbf w}_{\rm F}/2$
in q.~\eqref{Ztot}.  Each operator ${\cal O}$, appearing as
$\det({\cal O})$ in Eq.~\eqref{Ztot}, is of a Laplacian type and
unbounded; hence each determinant is infinite, rendering the whole
partition function ill-defined. The idea of spectral
  regularisation~\cite{Andrianov:1983fg} is to count eigenvalues of each operator that are
smaller than a cutoff scale $\Lambda$. If in addition one considers
the Euclidean space-time to have finite volume, then the spectrum of
each Laplacian becomes discrete and after its truncation one obtains a
product of finite number of modes.  Hence,
\be \det{\cal O} = \prod{\lambda_n} \xrightarrow[{{\scriptstyle
      \mathbf{spectral~regularisation}}}]{} \det\left( \frac{{\cal
    O}_{\Lambda}}{\mu^2}\right)= \prod_{\lambda_n
  \leq{\Lambda^2}}\frac{\lambda_n}{\mu^2}~, \ee
where
\be {\cal O}_{\Lambda}\equiv {\cal O} \cdot P_{\Lambda}~,\ee
with
 \be P_{\Lambda} \equiv
\Theta\left(\Lambda^2 - {\cal O}\right)~, \ee 
the projector on the
subspace of eigenfunctions of ${\cal O}$ with eigenvalues smaller than
$\Lambda$. The parameter $\mu$ is introduced in order to have a dimensionless
expression under the sign of determinant and in what follows, we consider $\Lambda=\mu$; other choices
of $\mu$ will not affect substantially the regularisation scheme.\footnote{The case of an arbitrary choice of $\mu$ is discussed in
Ref.~\cite{Andrianov:2011bc} for the fermionic determinant and can be
easily generalised for scalar or vector fields.}

Although the procedure of spectral regularisation can be easily
understood and has nice properties, like preserving gauge invariance
and general covariance, technically it is not easy to handle 
(in contrast to the Schwinger's proper time formalism). 
Nevertheless, one can address both, induced quantum gravity and anomaly-induced
inflation, using spectral regularisation. Indeed, they are both related with
Weyl non-invariance of the effective quantum action (or Weyl anomaly), since 
the classical theory is Weyl invariant. In the following, we compute Weyl anomaly and present the anomaly generating functional. 

\subsection{Spectral regularisation: computation of the Weyl anomaly}
Let us consider a conformal transformation of the metric tensor:
\be
g_{\mu\nu} \rightarrow \left(\widetilde{g_{\mu\nu} }\right)_{\phi} = e^{2\phi} g_{\mu\nu}~.
\ee
Since the classical action Eq.~\eqref{clas-act} is Weyl invariant, the
Weyl non-invariant contribution comes out, by definition, from Weyl anomaly. Let us compute the difference between the initial and the Weyl transformed quantum effective action, namely
\be
 W  - \widetilde{\left(W\right)}_{\phi} = \log\left(\frac{\widetilde{\left(Z\right)}_{\phi}} {Z}\right)~.
\label{diff}
\ee
For the fermionic effective action $W_{\rm F}$, this difference, Eq.~(\ref{diff}), reads~\cite{Andrianov:2011bc,Kurkov:2012dn}
\be \la{ScollF} W_{\rm F}  - \widetilde{\left(W_{\rm F}\right)}_{\phi} = - \int_0^1 {\rm d}t
\Tr
{\left(\phi\left[\widetilde{\chi\left(
\frac{\slashed{D}^2}{\Lambda^2}\right)}\right]_{\phi\cdot
t}\right)}~, \ee
where
\be \chi(z)\equiv\Theta(1-z)~.\la{chi}
\ee
Repeating the same computation for the case of a scalar field, one can
easily show that
\be \la{ScollH} W_{\mathcal{H}}  - \widetilde{\left(W_{\mathcal{H}}\right)}_{\phi} = \int_0^1 {\rm d}t
\Tr
{\left(\phi\left[\widetilde{\chi\left(
\frac{D_{ \mathcal{H}} }{\Lambda^2}\right)}\right]_{\phi\cdot
t}\right)}~. \ee
Indeed, under conformal
transformation the Laplacian $D_{ \mathcal{H}} $ transforms as
\be D_{ \mathcal{H}} \rightarrow\widetilde{\left(D_{ \mathcal{H}}
  \right)}_{\phi}\equiv e^{-3\phi}D_{ \mathcal{H}} e^{\phi}~, \la{hom}
\ee
and thus one obtains 
\ba W_{\mathcal{H}}  - \widetilde{\left(W_{\mathcal{H}}\right)}_{\phi} &=&
\log\left(\frac{\widetilde{\left(Z_{ \mathcal{H}}\right)}_{\phi}} {Z_{
    \mathcal{H}} }\right)\nonumber\\  &=& \int _{0}^1 {\rm d}t
\,\partial_t \log \widetilde{\left(Z_{
    \mathcal{H}}\right)}_{\phi(x)\cdot t}\no \nonumber &=& -\frac{1}{2}
\int_0^1 {\rm d}t \,\partial_t \Tr \left\{\log \left(\frac{\widetilde{D_{
      \mathcal{H}} }}{\Lambda^2}
\widetilde{P_{\Lambda}}\right)_{\phi(x)\cdot t}\right\} \no &=& -\frac{1}{2}
\int_0^1 {\rm d}t \,\Tr\left\{ \widetilde{D_{ \mathcal{H}}
}^{-1}\left(-3\phi \widetilde{D_{ \mathcal{H}} } + \widetilde{D_{
    \mathcal{H}} }\phi\right)\widetilde{P_{\Lambda}} +\underbrace{
  \left(\partial_t \Theta\left[\Lambda^2 - \widetilde{D_{ \mathcal{H}}
    }\right]\right)\cdot \log\frac{\widetilde {D_{ \mathcal{H}}
  }}{\Lambda^2}}_0 \right\}_{\phi\cdot t} \no &=& \int_0^1 {\rm d}t
\Tr \left\{\phi\widetilde{P_{\Lambda}}\right\}_{\phi\cdot t}~,
\la{comput} \ea
with
\be 
P_{\Lambda}\equiv \Theta\left(\Lambda^2 - D_{ \mathcal{H}} \right)~. 
\ee
Since the Laplacians $D_{\mathbf{vec}}$ and $D_{\mathbf{gh}}$ do not
transform in a homogeneous way, like $D_{\mathcal{H}}$ (see,
Eq.~\eqref{hom}), one cannot write a straightforward generalisation of
Eq.~\eqref{comput} for $D_{\mathbf{vec}}$ and
$D_{\mathbf{gh}}$. Nevertheless, there is a non-trivial interplay
between gauge and ghost modes and using the computation presented in
the Appendix one can generalise Eqs.~\eqref{ScollF}, \eqref{ScollH}. Hence,
defining 
\be
   W_{\mathbf{gauge}} \equiv W_{\mathbf{vec}} + W_{\mathbf{gh}}~,
\ee
one obtains 
\ba && W_{\mathbf{gauge}} - \widetilde{\left(W_{\mathbf{gauge}}\right)}_{\phi} = \no
&& \int_0^1 {\rm d}t
\left\{\Tr{\left(\phi\left[\widetilde{\chi\left(
      \frac{D_{\mathbf{vec}}}{\Lambda^2}\right)}\right]_{\phi\cdot
    t}\right)} -2
\Tr{\left(\phi\left[\widetilde{\chi\left(
      \frac{D_{\mathbf{gh}}}{\Lambda^2}\right)}\right]_{\phi\cdot 
    t}\right)} \right\}~. \la{ScollV}
\ea
To complete the computation of the scalar and gauge contributions to
the anomaly, Eqs.~\eqref{ScollH} and \eqref{ScollV}
respectively, we will follow the same procedure as in
Refs.~\cite{Andrianov:1985ay, Andrianov:1987jh, Vassilevich:1987yn,
  Andrianov:2011bc, Kurkov:2012dn}.

Let us first perform a decomposition of the projector $P_{\Lambda}$:
\bea P_\Lambda &=& \Theta\left(\Lambda^2 - {\cal O}\right)\nonumber\\
&=& \lim_{\epsilon\rightarrow 0}\,\frac{1}{2\pi i}
\int_{-\infty}^{+\infty}\frac{{\rm d}s}{s-i\epsilon}e^{is}
e^{-\left(\frac{is}{\Lambda^2}\right) {\cal O}}~,  \eea
and then do a heat kernel expansion\footnote{More precisely a
  Schr\"odinger kernel expansion, since the argument $z$ is purely
  imaginary.} in terms of the heat kernel (Schwinger-De Witt)
coefficients~\cite{Vassilevich:2003xt}:
\be \Tr \left(\phi \,e^{-z {\cal O}}\right) \simeq
\sum_{n=0}^{\infty}z^{\frac{1}{2}(n-4)}a_{n}\left(\phi,{\cal
O}\right)~, \ee
where 
\be
z = \frac{is}{\Lambda^2}~,
\ee
and
\be a_n\left(\phi,{\cal O}\right) = \int d^4 x\,\sqrt{g}\,\phi
\,a_n\left( {\cal O},x\right)~. \la{hc} \ee
The main advantage of the heat kernel method\footnote{Note that the
  heat kernel formalism is not valid beyond the one-loop
  approximation.} is that it provides the required information in
terms of only a few geometric invariants.  Since the ultraviolet
divergences in the one-loop effective action are defined by the heat
kernel coefficients, the asymptotic expansion, Eq.~\eqref{hc}, makes
sense only when the background field invariants, appearing in the heat
kernel coefficients $a_n\left( {\cal O},x\right)$, are smaller than
the corresponding powers of the ultraviolet cutoff $\Lambda$.  This
requirement defines the applicability of our approach; we assume this
requirement to be satisfied.\footnote{In the case of anomaly-induced
  inflation, one must check that the scalar curvature $R$ is small
  enough with respect to the ultraviolet cutoff scale $\Lambda^2$;
  this is indeed the case.}  Since we are working on a
manifold without boundary, only even heat kernel coefficients $a_{2k}$
are non-zero.

Performing the integration over $s$ in Eq.~\eqref{hc}, namely
\be \int_{-\infty}^{+\infty} {\rm d}s~s^{k-3}e^{is} =
\left\{\begin{array}{ll} \frac{1}{2\pi
  i}~\mbox{Res}_{s=0}~\left(s^{k-2}e^{is}\right)~~~~\mbox{for}~~~~k =
0, 1, 2 ~;
\\ \left(2\pi\right)i^{k-3}\left(\partial^{(k-3)}\delta\right)(1)
=0~~~~\mbox {for}~~~~k\geq 3~,
\end{array}
\right. 
\ee
we obtain
\be \Tr\left( \phi ~\Theta\left(\Lambda^2 - {\cal O}\right) \right) =
\int {\rm d}^4 x \sqrt{g}\left(\frac{a_0({\cal O},x)}{2} \Lambda^4 +
a_2({\cal O},x) \Lambda^2 + a_4({\cal O},x) \right) ~.\la{expansion}
\ee
Using the expansion Eq.~\eqref{expansion} for the total anomaly we obtain
\ba 
W - \widetilde{\left(W\right)}_{\phi} &=& \int {\rm d}^4x ~\phi(x)\int_0^{1} {\rm d}t
\sqrt{\tilde g_{\phi t}} \left\{\frac{\Lambda^4}{2} \left(N_{\mathcal
  H} a_0^{\mathcal{H}} + N_V\left[a_0^{\mathbf{vec}} - 2
  a_0^{\mathbf{gh}}\right] - \frac{N_{\rm F}^{\mathbf{w}}}{2}
a_0^{{\rm F}}\right)\right.\nonumber \\
&&~~~ +\Lambda^2\left(N_{\mathcal H}
\widetilde{\left(a_2^{\mathcal{H}}\right)}_{\phi\cdot t} +
N_V\left[\widetilde{\left(a_2^{\mathbf{\mathbf{vec}}}\right)}_{\phi\cdot
    t} - 2
  \widetilde{\left(a_2^{\mathbf{\mathbf{gh}}}\right)}_{\phi\cdot
    t}\right] - \frac{N_{\rm F}^{\mathbf{w}}}{2}
\widetilde{\left(a_2^{{\rm F}}\right)}_{\phi\cdot t}\right) \nonumber \\
&&~~~ +
\left. \left(N_{\mathcal H}
\widetilde{\left(a_4^{\mathcal{H}}\right)}_{\phi\cdot t} + N_{\rm
  V}\left[\widetilde{\left(a_4^{\mathbf{\mathbf{vec}}}\right)}_{\phi\cdot
    t} - 2
  \widetilde{\left(a_4^{\mathbf{\mathbf{gh}}}\right)}_{\phi\cdot
    t}\right] - \frac{N_{\rm F}^{\mathbf{w}}}{2}
\widetilde{\left(a_4^{{\rm F}}\right)}_{\phi\cdot t}\right)
\right\}.\label{Scollcomput} \ea
We give in Tables~\ref{tab:a02} and \ref{tab:a4} the values of the
heat kernel coefficients $a_0$, $a_2$ and $a_4$, respectively, for
free massless fields of different spin. Note that $C^2=
C_{\mu\nu\rho\sigma}C^{\mu\nu\rho\sigma}$ with $C_{\mu\nu\rho\sigma}$
the Weyl tensor, and {\bf GB} stands for the Gauss-Bonnet term, given
by
{\bf GB}$=C^2
-2\left[R_{\mu\nu}R^{\mu\nu}-(1/3)R^2\right].$
\begin{table}
\caption{\label{tab:a02} Heat kernel coefficients $a_0$ and $a_2$ for
free massless fields of various spin; we have calculated them using Ref.~\cite{Vassilevich:2003xt}.}
\begin{center}
\begin{tabular}{|c|c|c|}
\hline Spin & $a_0$ & $a_2$\\ \hline 0, conformal coupling &
$\frac{1}{16\pi^2}\cdot 1$ & 0 \\ \hline 1/2, Dirac fermion &
$\frac{1}{16\pi^2}\cdot 4$ &
$\frac{1}{16\pi^2}\left(\frac{R}{3}\right)$\\ \hline 1, without ghosts
& $\frac{1}{16\pi^2}\cdot 4$ &
$\frac{1}{16\pi^2}\left(\frac{R}{3}\right)$\\ \hline 0, minimal
coupling & $\frac{1}{16\pi^2}\cdot 1$ &
$-\frac{1}{16\pi^2}\left(\frac{R}{6}\right)$\\ \hline 1, gauge
(i.e., with ghosts) & $\frac{1}{16\pi^2}\cdot 2$ &
$\frac{1}{16\pi^2}\left(\frac{2 R}{3}\right)$\\ \hline
\end{tabular}
\end{center}
\end{table}
\begin{table}
\caption{\label{tab:a4} Heat kernel coefficient $a_4$ for free
massless fields of various spin~\cite{Vassilevich:2003xt}.}
\begin{center}
\begin{tabular}{|c|c|c|c|}
\hline \multicolumn{4}{|c|}{$a_4 = \frac{1}{2880\pi^2}\left(a\cdot C^2
+ b\cdot \mathbf{GB} + c \cdot R_{;\mu}^{~\mu}\right)$} \\ \hline Spin
& $a$ & $b$ &$c$\\ \hline 0, conformal coupling & 3/2 & -1/2 & -1 \\
\hline 1/2, Dirac fermion & -9 & 11/2 & 6 \\ \hline 1, gauge
(i.e., with ghosts) & 18 & -31 & 18\\ \hline
\end{tabular}
\end{center}
\end{table}
\\
Using
\ba \int {\rm d}^4 x ~\phi(x) \int_0^1\widetilde{\left(\sqrt{g}
  \mathbf{GB}\right)}_{\phi\cdot t} &=& \int {\rm d}^4 x
\sqrt{g}\left(\phi~\mathbf{GB}+ 4 G^{\mu\nu}\phi_{\mu}\phi_{\nu} - 4XY
- 2X^2\right), \no \int d^4 x ~\phi(x)
\int_0^1\widetilde{\left(\sqrt{g} R_{;\mu}^{~\mu}\right)}_{\phi\cdot
  t} &=& \int {\rm d}^4 x \sqrt{g}\left((X+Y)R + 3(X+Y)^2\right), \no
\int {\rm d}^4 x ~\phi(x) \int_0^1\widetilde{\left(\sqrt{g}
  R\right)}_{\phi\cdot t} &=& \int {\rm d}^4 x
\sqrt{g}\left(\frac{1}{2}\left(\e^{2\phi} -1 \right)R - 3~e^{2\phi} X
\right)~,\label{formulas} \ea
where
\be \phi_{\mu}\equiv \partial_{\mu}\phi,\quad X\equiv
\phi_{\mu}\phi^{\mu},\quad Y\equiv \nabla^{\mu}\phi_{\mu}~.  \ee
the total anomaly Eq.~\eqref{Scollcomput} reads
\ba\la{Scollansw} W - \widetilde{\left(W\right)}_{\phi}  &=& \int {\rm d}^4 x
\sqrt{g}\left\{\alpha_1\left(e^{4\phi} -1\right)
+\alpha_2\left(\frac{1}{2}\left(\e^{2\phi} -1 \right)R - 3~e^{2\phi} X
\right) +\alpha_3\phi~C^2
\right. \nonumber\\ &&~~~~~~~~~~~~~+\alpha_4\left(\phi~\mathbf{GB}+ 4
G^{\mu\nu}\phi_{\mu}\phi_{\nu} - 4XY - 2X^2\right) \nonumber\\
&&~~~~~~~~~~~~~
+
\left.\alpha_5\left((X+Y)R + 3(X+Y)^2\right)\right\}~, \ea
where
\ba \alpha_1 &\equiv& \frac{\Lambda^4}{128\pi^2}\left(N_{\mathcal H} + 2
N_{\rm V} - 2 N_{\rm F}^{\mathbf{w}}\right),\no \alpha_2 &\equiv&
\frac{\Lambda^2}{16\pi^2}\left( - \frac{1}{6} N_{\rm F}^{\mathbf{w}} +
\frac{2}{3}N_{\rm V}\right),\no \alpha_3 &\equiv& \frac{1}{2880
  \pi^2}\left(\frac{3}{2}N_{\mathcal H} + \frac{9}{2}N_{\rm
  F}^{\mathbf{w}} + 18 N_{\rm V}\right),\no \alpha_4 &\equiv&
-\frac{1}{2880 \pi^2}\left(\frac{1}{2}N_{\mathcal H} +
\frac{11}{4}N_{\rm F}^{\mathbf{w}} + 31 N_{\rm V}\right), \no \alpha_5
&\equiv& \frac{1}{2880 \pi^2}\left(- N_{\mathcal H} - 3 N_{\rm
  F}^{\mathbf{w}} + 18 N_{\rm V} \right)~.  \la{coefs}\ea
At this point, one can make two remarks:
\\
\underline{\sl Remark 1}\\
We would like to compare results for the trace anomaly obtained via the spectral 
and the $\zeta$-function regularisations.
An infinitesimal anomaly reads
\be \la{infanom}
\lim_{\phi\rightarrow 0} ~\frac{1}{\sqrt{g}}\,\,{ \frac{\delta }{\delta \phi(x)}} \,\,\widetilde{\left(W\right)}_{\phi}
= - \left(\frac{\Lambda^4}{2} \cdot A_0(x) + \Lambda^2 \cdot A_2(x) + \Lambda^0 \cdot A_4(x)\right)~, 
\ee
where
\ba
 A_0 &\equiv& N_{\mathcal
  H} \,a_0^{\mathcal{H}} + N_V\,\left[a_0^{\mathbf{vec}} - 2
  a_0^{\mathbf{gh}}\right] - \frac{N_{\rm F}^{\mathbf{w}}}{2}\,
a_0^{{\rm F}}~,\no
A_2 &\equiv& N_{\mathcal H}\,
a_2^{\mathcal{H}} +
N_V\,\left[{a_2^{\mathbf{\mathbf{vec}}} - 2
  a_2^{\mathbf{\mathbf{gh}}}}\right] - \frac{N_{\rm F}^{\mathbf{w}}}{2} \,
{a_2^{{\rm F}}}~,\no
A_4 &\equiv& N_{\mathcal H}
{a_4^{\mathcal{H}}} + N_{\rm
  V}\left[{\left(a_4^{\mathbf{\mathbf{vec}}}\right)} - 2
  {a_4^{\mathbf{\mathbf{gh}}}}\right] - \frac{N_{\rm F}^{\mathbf{w}}}{2}
{a_4^{{\rm F}}}~,
\ea
and the heat kernel coefficients $a_0, a_2, a_4$  are given in Tables~\ref{tab:a02} and \ref{tab:a4}.
The $A_4$-contribution coincides 
with the result for anomaly obtained via $\zeta$-function regularisation and the dimensional one~\cite{Birrell:1982ix}.
Quadric and quadratic in $\Lambda$ terms can be interpreted as an ultraviolet divergence and hence subtracted  through the addition of the corresponding local counter terms.
Indeed, one can define the renormalised effective action 
\be
W^{\rm ren} \equiv W  +  \int {\rm d}^4 x \sqrt{g}
\left(\alpha_1+\alpha_2 \left(\frac{R}{2}\right)\right),
\ee
with $\alpha_1,\alpha_2$ defined in Eq.~\eqref{coefs}.
One can easily check  (see computations in subsection~\ref{subsec:indgravspec}) that
\be
\lim_{\phi\rightarrow 0} ~\frac{1}{\sqrt{g}}\,\,{ \frac{\delta }{\delta \phi(x)}} \,\,\widetilde{\left(W^{\rm ren}\right)}_{\phi}
= - A_4(x), \la{standanom}
\ee
with $A_4$ defined in Eq.~\eqref{infanom}.
However in this way, spectral regularisation does not lead to any new result. 

In what follows, we will not subtract the divergent terms and we will keep $\Lambda$ finite and of order of the Planck scale. We will thus be able to describe simultaneously both, the induced gravitational action and the onset of
(trace) anomaly-induced inflation. We will hence conclude that all terms in the Lagrangian, leading to a period of an accelerated expansion of the universe, may be considered as the outcome of a quantum effect.
\\
\underline{\sl Remark 2}\\
As we will show, fermions play an important r\^ole in both, induced gravity and anomaly-induced inflation, and hence the case of a purely fermonic anomaly is of special importance.
This was studied in Refs.~\cite{Andrianov:2010nr, Andrianov:2011bc, Kurkov:2012dn}, 
and the Weyl anomaly generating functional was expressed in terms of a structure very similar to the Chamseddine and Connes bosonic spectral action~\cite{Chamseddine:1996zu}, introduced
in a context of noncommutative spectral geometry~\cite{ncg-book1,ncg-book2} coupled to the collective dilaton.\footnote{Noncommutative spectral geometry provides a purely geometric explanation for the Standard Model of particle physics~\cite{ccm,cchiggs,Chamseddine:2013rta,Devastato:2013oqa} and offers a framework to address early universe cosmology~\cite{Buck:2010sv,Nelson:2008uy,Nelson:2009wr,Marcolli:2009in,Nelson:2010ru,Nelson:2010rt,Lambiase:2013dai}.}

At this point, let us emphasise that the infinitesimal Weyl anomaly, obtained within QFT with spectral regularisation, is the bosonic spectral Lagrangian. Indeed, by definition,
\ba
S_{\rm BS} &\equiv& \Tr\left({\chi\left(\frac{\slashed{D}^2}{\Lambda^2}\right)}\right)\nonumber\\
&\simeq& \int \,d^4x \,\sqrt{g}\, L_{\rm BS}(x), 
\ea
where $\chi(z)$ is a cutoff function, equal to one at $z<1$ and rapidly vanishing at $z > 1$, and $L_{\rm BS}(x)$ stands for
the bosonic spectral  Lagrangian, computed via the heat-kernel technique. Performing a similar computation and inserting $\phi(x)$ under the sign of the trace, we get
\be
\Tr\left(\phi\,\left[\chi\left(\frac{\slashed{D}^2}{\Lambda^2}\right)\right]\right)
\simeq\int \,d^4x\, \sqrt{g}\, \phi\, L_{\rm BS}(x)~, 
\ee
and the bosonic spectral Lagrangian reads
\be \la{BSL}
L_{\rm BS}(x) = \frac{1}{\sqrt{g}} \,\,\frac{\delta }{\delta \phi(x)}\,\, \Tr\left(\phi\, \left[\chi\left(\frac{\slashed{D}^2}{\Lambda^2}\right)\right]\right)~.
\ee
Expanding Eq.~\eqref{ScollF} up to linear order in $\phi$
and taking the functional derivative in the infinitesimal limit, we obtain:
\ba
\mbox{infinitesimal Weyl anomaly}&\equiv&\lim_{\phi\rightarrow 0} ~\frac{1}{\sqrt{g}}\,\,{ \frac{\delta }{\delta \phi(x)}}
\,\, \widetilde{\left(W_{\rm{F}}\right)}_{\phi} \no 
&=&\lim_{\phi\rightarrow 0}\,\,\frac{1}{\sqrt{g}} \,\,\frac{\delta }{\delta \phi(x)}\,\,\int_0^1 {\rm d}t
\Tr{\left(\phi\left[\widetilde{\chi\left(
\frac{\slashed{D}^2}{\Lambda^2}\right)}\right]_{\phi\cdot
t}\right)} \no
&=&\lim_{\phi\rightarrow 0}\,\,\frac{1}{\sqrt{g}} \,\, \frac{\delta }{\delta \phi(x)}\,\, 
\left[\Tr\left(\phi\, \left[\chi\left(\frac{\slashed{D}^2}{\Lambda^2}\right)\right]\right)
 + O\left(\phi^2\right)\right] \no
 & = & \frac{1}{\sqrt{g}}\,\, \frac{\delta }{\delta \phi(x)} \,\,\Tr\left(\phi\, \left[\chi\left(\frac{\slashed{D}^2}{\Lambda^2}\right)\right]\right)\no
 & = & L_{\rm BS}(x)~,\la{anomBSL}
\ea
where in the last step we used Eq.~\eqref{BSL}.

\subsection{Anomaly generating functional and collective dilaton}
Although -- in contrast to proper time regularisation -- spectral regularisation does not allow one to compute the partition function explicitly,  there is  a formalism via the introduction of {\sl a collective dilaton} that allows one to extract the Weyl non-invariant part of such a regularised determinant, as an integral over an auxilary field $\phi$ of some local expression that depends on $\phi$ and the background fields.

Substituting the conformally transformed metric tensor
$g_{\mu\nu}e^{2\phi}$ in Eq.~\eqref{Ztot} and integrating over all
possible $\phi(x)$, one can write the identity
\be
\label{Z-fun}
Z = \left(\int [{\rm d}\phi]
\widetilde{\left(Z^{-1}\right)}_{\phi}\right)^{-1} \cdot \int [{\rm
    d}\phi]\left(\frac{Z}{\widetilde{\left(Z\right)}_{\phi}}\right)~.
\ee
Since the first term above is the integral over the Weyl group of a
Weyl transformed quantity, it is Weyl invariant under the action of
the Weyl group, so we denote it by $Z_{\mathbf{inv}}$.  Hence,
Eq.~(\ref{Z-fun}) can be rewritten as
\be\label{part-f}
Z \equiv Z_{\mathbf{inv}}\cdot \int [{\rm d}\phi] e^{-S_{\mathbf{coll}}}~,
\ee
where
\be  S_{\mathbf{coll}}
\equiv \log\left(\frac{\widetilde{\left(Z\right)}_{\phi}}{Z}\right)~.
\ee
Thus, the non-Weyl invariant partition function $Z$ in
Eq.~(\ref{part-f}) is written as the product of a term $Z_{\rm inv}$
invariant under Weyl transformations and another one, non-invariant,
which depends on the auxiliary field $\phi$ and is due to Weyl
anomaly. The introduction of the auxiliary field, representing the
collective degree of freedom of all fermions, can be seen as {\sl
  bononisation}.  As we will later show, there exists a {\sl local}
Lagrangian $L_{\mathbf{coll}}$ depending on $\phi$ and background
fields, such that $S_{\mathbf{coll}} = \int d^4 x\sqrt{g}L$.  Hence,
instead of computing $Z$, we can use a bosonisation-like relation
\bea Z[g_{\mu\nu}] &=& \int[{\rm d}\bar \psi][{\rm d}\psi][{\rm d}
  \mathcal{H}][{\rm d} A][{\rm d}\bar c][{\rm d} c]
e^{-S_{\mathbf{cl}}[\bar\psi,\psi,\mathcal{H},A,\bar c, c,g_{\mu\nu}]}
\\
\nonumber &\simeq& Z_{\mathbf{inv}}\cdot
\int[D\phi]e^{-S_{\mathbf{coll}}[\phi, g_{\mu\nu}]}~.  \eea
Clearly, $\phi$ stands for a collective degree of freedom of vacuum
fluctuations of all fields dual to conformal anomaly, hence the term
``collective action".

Since all our computations were carried in Euclidean QFT, in
order to apply our result in a physical context  one
should perform a Wick rotation back to Minkowski signature in
Eq.~\eqref{Scollansw}. Hence, for the anomaly generating functional we have 
\be Z_{\mathbf{coll}}\equiv \int [d\phi] e^{-S_{\mathbf{coll}}}
\xrightarrow[{{\scriptstyle \mbox{Wick rotation back}}}]{}
Z_{\mathbf{coll}~\mathbf{M}}\equiv \int[d\phi] e^{i
  S_{\mathbf{coll}~\mathbf{M}}}~, \ee
and the Minkowskian version of
the collective action reads 
\ba S_{\mathbf{coll}~\mathbf{M}} &=& \int
{\rm d}^4 x \sqrt{-g}\left(-\alpha_1\left(e^{4\phi} -1\right)
+\alpha_2\left(\frac{1}{2}\left(\e^{2\phi} -1 \right)R - 3~e^{2\phi} X
\right) -\alpha_3\phi~C^2 \right. \nonumber\\ &&
~~~~~~~~~~~~~~~~~ - \alpha_4\left(\phi~\mathbf{GB}+ 4
G^{\mu\nu}\phi_{\mu}\phi_{\nu} - 4XY - 2X^2\right)\nonumber\\
&&~~~~~~~~~~~~~~~~~ -
\left.\alpha_5\left((X+Y)R + 3(X+Y)^2\right)\right)~, \la{Minkansw} \ea 
with
the coefficients given in Eq.~\eqref{coefs}.

In what follows, we will show that Weyl anomaly in QFT with spectral regularisation 
reproduces Sakharov's induced gravity, as well as Starobinsky's anomaly-induced inflation. This is the main message of our study.

\section{Sakharov's Induced Gravity and Spectral Regularisation
\la{sec:indgrav}}
\subsection{Standard Approach: Proper Time Regularisation}
The standard approach to the Sakharov's induced gravity is based on
Schwinger's proper time regularisation~\cite{Visser:2002ew}. In this
formalism, one first selects a convenient reference metric $\tilde
g_{\mu\nu}$ and then computes the difference in the one-loop
contribution to the effective action which results from comparing two
different metrics defined on the same topological manifold. 
Hence, we
consider the difference $W[g_{\mu\nu}] - W[\tilde{g_{\mu\nu}}] $,
with $W$ defined as $W\equiv -\log Z$.  

Let us write the formal equality
\bea \Tr\left( \log{\frac{D}{\tilde D}}\right) &=&
\sum_{n=0}^{\infty}\log{\frac{\lambda_n}{\tilde\lambda_n}} \no\\ &=&
-\sum_{n=0}^{\infty}\int_0^{\infty} {\rm d} s
\left(\frac{e^{-s\lambda_n}}{s} -
\frac{e^{-s\tilde\lambda_n}}{s}\right) \no\\ &=&
-\int_0^{\infty}\frac{{\rm d}s}{s} \Tr\left(e^{-sD} - e^{s\tilde D}\right)~,
\eea
and then perform a heat kernel expansion for the $\Tr (e^{-sD})$ and
$\Tr(e^{s\tilde D})$ terms to get
\be \Tr\left( \log{\frac{D}{\tilde D}}\right) = -\int_0^{\infty}\frac{{\rm d}s}{s}
\sum_{k=0}^{\infty}s^{k-2}\left(a_{2k}(D) - a_{2k}(\tilde D)\right)~,
\la{hcform} \ee
where the coefficients $a_k$ are the Seeley-De Witt coefficients,
universal functions of the space-time geometry.  In order to perform
the integration over $s$ in Eq.~\eqref{hcform} for $k=0, 1$ one needs
an ultraviolet regulator $\mu_{\mathbf{uv}}$; integration over $s$ for
all other values of $k$, namely for all $k>1$, is ultraviolet finite
but it requires the infrared regulator $\mu_{\mathbf{ir}} \ll
\mu_{\mathbf{uv}}$. It is worth noting that the heat kernel expansion
has allowed us to identify the potential divergences.\\  
We obtain
\ba \Tr \left( \log{\frac{D}{\tilde D}} \right)&=&
-\int_{\mu_{\mathbf{uv}}^{-2}}^{\mu_{\mathbf{ir}}^{-2}} \frac{{\rm d}s}{s}
\sum_{k=0}^{\infty}s^{k-2}\left(a_{2k}(D) - a_{2k}(\tilde
D)\right) \no \no\ &=&
-\frac{\mu_{\mathbf{uv}}^4}{2}\left(a_0\left(D\right)
-a_0\left(\tilde{D}\right)\right)
-\mu_{\mathbf{uv}}^2\left(a_2\left(D\right)
-a_2\left(\tilde{D}\right)\right)\no\no
&&-\log\left(\frac{\mu_{\mathbf{uv}}^2}{\mu_{\mathbf{ir}}^2}\right)
\left(a_4\left(D\right)
- a_4\left(\tilde{D}\right)\right) + \cdots \la{expand} \ea
Let us emphasise that the regulators $\mu_{\mathbf{uv}}$ and
$\mu_{\mathbf{ir}}$ are not ultaviolet and infrared, respectively,
cutoff scales for the spectrum of $D$; they are attributes to make the
regularisation scheme finite.\footnote{Considering $D = -\partial^2 +
m^2$ or $D = -\partial^2$ and a finite volume of Euclidean spacetime,
the spectrum D has an infrared cutoff, however the integration over
$s$ in Eq.~\eqref{hcform} is still infrared divergent.} \\ Using
Eq.~\eqref{Ztot} with $\log\det =\Tr\log$ and Eq.~\eqref{expand} we
get
\be W^{\mathbf{pt}}\equiv -\log Z =  \int d^4 x\sqrt{g}\left(\lambda_{\mathbf{ind}}^{\mathbf{pt}} +
\frac{M_{\rm Pl}^{2~\mathbf{ind}}}{16\pi}R +
\left\{O\left(R^2\right)\right\}\right)~, \la{WeffPT}\ee 
where
\be \lambda_{\mathbf{ind}}^{\mathbf{pt}}
=\frac{\mu_{\mathbf{uv}}^4}{64\pi^2}\left(2N_{\rm F}^{\mathbf{w}} - N_{\mathcal H}
-2N_{\rm V}\right)~, \ee
and
\be M_{\rm
Pl}^{2~\mathbf{ind}}=\frac{\mu_{\mathbf{uv}}^2}{2\pi}\left(
\frac{N_{\rm F}^{\mathbf{w}}}{6}
-\frac{2N_{\rm V}}{3}\right)~.  \ee
The main idea of Sakharov's induced gravity lies in attributing a
physical meaning to the ultraviolet cutoff scale, so that it denotes
the upper scale for which the considered QFT is a valid effective
theory.  In this way, it is not necessary to subtract divergences, and
setting $\Lambda\sim M_{\rm Pl}\sim 10^{19}$GeV, the term
$\mu_{\mathbf{uv}}^2 R$ can be considered as an induced gravitational
action.

Hence, starting from a classically Weyl invariant theory, quantisation
implied a Weyl non-invariant Einstein-Hilbert action. One may thus
conclude that, under proper time regularisation, the Weyl anomaly 
contains operators of dimension two, in contrast to the (standard) dimensional regularisation or the $\zeta$-function regularisation, where anomaly contains just
operators of dimension four.

Nevertheless, the considered {\sl  proper time} regularisation procedure does not reproduce correctly the $a_4$-contribution to the anomaly (c.f. Eq.~\eqref{standanom}), and
therefore it cannot be used to investigate the trace anomaly induced
inflation.  Indeed, substituting in $W^{\mathbf{pt}}$, defined in \eqref{WeffPT} above, the conformally transformed metric tensor $e^{2\phi}g_{\mu\nu}$  and then taking the derivative over $\phi(x)$,
one immediately finds 
\be
\lim_{\phi\rightarrow 0} ~\frac{1}{\sqrt{g}}\,\,{ \frac{\delta }{\delta \phi(x)}} \,\,\widetilde{\left(W^{\mathbf{pt}}\right)}_{\phi} = 4~\lambda_{\mathbf{ind}}^{\mathbf{pt}} + \frac{1}{8\pi}M_{\rm
Pl}^{2~\mathbf{ind}} R~, \la{ptanom}
\ee
taking into account that the $\left\{O\left(R^2\right)\right\}$-terms  in Eq.~\eqref{WeffPT} are  given by
\ba \log\left(\frac{\mu_{\mathbf{uv}}^2}{\mu_{\mathbf{ir}}^2}\right)
a_4\left(D\right)
& =& \log\left(\frac{\mu_{\mathbf{uv}}^2}{\mu_{\mathbf{ir}}^2}\right)
\frac{1}{2880\pi^2}\left[\frac{3}{2}N_{\mathcal H} + \frac{9}{2}N_{\rm
  F}^{\mathbf{w}} + 18 N_{\rm V}\right]
\int d^4 x\sqrt{g} C^2 \nonumber\\
&=& \mbox{\bf Weyl inv.} ~, \la{logconrtib}
\ea 
and thus do not contribute in Eq.~\eqref{ptanom}.

Let us remind to the reader that as we have previously shown (see Remark 1,  Eq.~\eqref{infanom}), the spectral regularisation reproduces correctly the $a_4$-contribution to the anomaly.
We will next show that it also reproduces
correctly the induced Einstein-Hilbert action; it can be thus used to describe
both.

\subsection{The Spectral Regularisation Approach}
\la{subsec:indgravspec}
The effective action  
\be
W_{\mathbf{eff}}[g_{\mu\nu}] = -\log{Z[g_{\mu\nu}]}~,
\ee
is known to be a non-local functional of the metric tensor $g_{\mu\nu}$ and in particular, of the Lagrangian density $L_{\mathbf{eff}}[g_{\mu\nu}]$, so that
\be W_{\mathbf{eff}}[g_{\mu\nu}] = \int d^4 x
\sqrt{g}L_{\mathbf{eff}}[g_{\mu\nu}] \ee
does not exist and correspondingly the local collective action
$S_{\mathbf{coll}}$, once integrated over $\phi$, captures all
non-locality of the Weyl non-invariant part of the effective action.
 
Nevertheless, the terms with coefficients $\alpha_1$, $\alpha_2$ and
$\alpha_5$ in the anomaly, Eq.~\eqref{Scollansw}, can
be generated by local terms in the effective action $W_{\mathbf{eff}}$.  \\ 
Indeed, let us consider
\be
 W_{\mathbf{loc} }[g_{\mu\nu}] = \int {\rm d}^4 x \sqrt{g}
\left(-\alpha_1-\alpha_2 \left(\frac{R}{2}\right) -\alpha_5
\left(\frac{R^2}{12}\right)\right)~,\label{Wloc} \ee
and
\ba 
\la{Scollanswloc} W_{\mathbf{loc}}[g_{\mu\nu}] - W_{\mathbf{loc}}[g_{\mu\nu}e^{2\phi}]&=& \int {\rm d}^4 x \sqrt{g}\left(\alpha_1\left(e^{4\phi} -1\right)
+\alpha_2\left(\frac{1}{2}\left(\e^{2\phi} -1 \right)R - 3~e^{2\phi} X
\right)\right.\no && \left.\ \ \ \ \ \ \ \ \ \ \ \ \ \ \
+\alpha_5\left((X+Y)R + 3(X+Y)^2\right)\right)~.  \ea
Comparing Eqs.~\eqref{Scollansw}\ and \eqref{Scollanswloc},
we conclude that the total effective action can be rewritten as
\be W = W_{\mathbf{inv}} + W_{\mathbf{loc}} +
W_{\mathbf{nonloc}}~, \ee
where $W_{\mathbf{inv}}$ is some Weyl invariant functional of
$g_{\mu\nu}$ and $W_{\mathbf{nonloc}}$ is a non-local functional,
generating $\alpha_3$ and $\alpha_4$ terms in the collective action
Eq.~\eqref{Scollansw} (we refer the reader to Ref.~\cite{Riegert:1984kt}).  Equivalently, one can say that QFT with spectral regularisation leads to Sakharov's induced gravity:
\ba W_{\mathbf{ind}}\left[g_{\mu\nu}\right] &=& \int {\rm d}^4 x
\sqrt{g}\left( \frac{\Lambda^4}{128\pi^2}\left(2 N_{\rm
  F}^{\mathbf{w}}-N_{\mathcal H} - 2 N_{\rm V} \right)
+\frac{\Lambda^2}{32\pi^2}\left(\frac{1}{6} N_{\rm
  F}^{\mathbf{w}} - \frac{2}{3}N_{\rm V}\right)R +
O\left(\left\{R^2\right\}\right)\right)\no &=& \int {\rm d}^4 x
\sqrt{g}\left(\lambda^{\mathbf{ind}} + \frac{1}{16\pi} \left(M_{\rm
  Pl}^{\mathbf{ind}}\right)^2 R \right) +
O\left(\left\{R^2\right\}\right)~, \ea
where
\be \left(M_{\rm Pl}^{\mathbf{ind}}\right)^2 =
\frac{\Lambda^2}{12\pi}\left(N_{\rm F}^{\mathbf{w}} - 4N_{\rm
  V}\right)~, \ee
and
\be \lambda^{\mathbf{ind}} = \frac{\Lambda^4}{128\pi^2}\left(2
N_{\rm F}^{\mathbf{w}}-N_{\mathcal H} - 2 N_{\rm V} \right)~, \ee 
with $O\left(\left\{R^2\right\}\right)$ denoting all local and
non-local terms responsible for $\Lambda^0$-contributions in the
anomaly-induced effective action. The latter is much smaller in the
low energy regime ($R<<\Lambda$) and hence it can be neglected at
energies much smaller than the cutoff scale. It however plays a
significant r\^ole during the inflationary era; it will be studied in
the next section within the isotropic approximation.

In order to identify the induced Planck mass with the real one at
$\sim10^{19}$GeV one should impose the cutoff scale $\Lambda$ at the
Planck energy scale. This however automatically leads to a huge value
of the induced cosmological constant, namely
$\lambda^{\mathbf{ind}}\sim M_{\rm Pl}^4$. One may consider the
presence of bare cosmological constant with the opposite sign, namely
$\lambda^{\mathbf{bare}}\sim -~M_{\rm Pl}^4$ and impose the
fine-turning:
\be \lambda^{\mathbf{observable}} = \lambda^{\mathbf{bare}} +
\lambda^{\mathbf{ind}}~.\no \ee
To avoid such a fine-tuning, we will adopt an alternative approach and hence, we will impose the Pauli compensation principle, i.e we require, that the $\Lambda^4$ fermonic and bosonic contributions to the vacuum energy cancel each other. The latter implies that the numbers of physical fermonic 
and bosonic degrees of freedom are equal, namely
\be N_{\mathcal H} = 2\left(N_{\rm F}^{\mathbf{w}} - N_{\rm
  V}\right)~,\la{Weltman} \ee
on the number of scalars, spinors and vectors, so that all quartic divergences cancel.
\\
Thus, under spectral regularisation we obtain:
\be W_{\mathbf{ind}}\left[g_{\mu\nu}\right] = \int {\rm d}^4 x
 \sqrt{g}\left(\frac{\Lambda^2}{32\pi^2}\left(\frac{1}{6}
 N_{\rm F}^{\mathbf{w}} - \frac{2}{3}N_{\rm V}\right)R +
 O\left(\left\{R^2\right\}\right)\right)~. \la{Sind} \ee
The above equation, Eq.~\eqref{Sind}, agrees with the one obtained
following the Schwinger proper time formalism~\cite{Visser:2002ew}.
It is worth noting that the Pauli compensation condition $N_{\mathcal H} =
2\left(N_{\rm F}^{\mathbf{w}} - N_{\rm V}\right)$ is not just a
property of the spectral regularisation; it holds in all
regularisation procedures with an ultraviolet cutoff scale and in that
sense it is universal.

In the next section, we will consider a high energy region, but $R<\Lambda^2$, i.e. where the spectral regularisation is still applicable. We will show that, imposing the Pauli compensation condition, the $\Lambda^0$-contribution to the anomaly (which we have neglected here), together with the $\Lambda^2$-contribution, leads automatically to Starobinsky's anomaly-induced inflation.

\section{Inflation Induced from Trace Anomaly: The Isotropic Approximation 
\la{sec:infl}}
We will explore the dynamics of a metric tensor in the isotropic
approximation.  The spacetime is considered to be spatially flat,
namely $g_{\mu\nu} = e^{\beta(\tau)}\eta_{\mu\nu}$; the cases of
closed and open universes can be studied along similar lines.
Although one should first derive an equation of motion for the metric
tensor $g_{\mu\nu}$ and only afterwards substitute the conformally
flat anzatz, it is possible to avoid the first step following the
procedure described in Ref.~\cite{Weinberg:2009wa}.

Hence, to get equations of motion in the isotropic case for an arbitrary\footnote{Although the procedure discussed is Ref.~\cite{Weinberg:2009wa} deals with a local action $W$,
repeating the same analysis one obtains the same result also in a
nonlocal situation, provided after the substitution of the
conformally flat anzatz, the action can be written as the
right-hand-side of Eq.~\eqref{Form}. As we will see, our model
belongs to this case.}  general covariant action $W[g_{\mu\nu}]$
 one should~\cite{Weinberg:2009wa}:
\begin{itemize}
\item Firstly, substitute the conformally flat anzatz ${\rm d}s^2 =
  {\rm d}t^2 - a(t)^2 {\rm d}\vec{x}^2$ in the action $W[g_{\mu\nu}]$.
\item Secondly, rewrite the result of the substitution in the form
\be
W[a(t)] = {\rm{vol}}\cdot \int~ {\rm d} t ~a^3 ~{\mathfrak I}(H,\dot H), \la{Form}
\ee
where $\rm{vol}$ is a three-dimensional volume and $H$ stands for the
Hubble parameter, $H\equiv \dot a/ a$. In principle, the function
${\mathfrak I}$ can also depend on higher derivatives of the Hubble
parameter, but we will restrict ourselves to the minimal needed case.
\item Thirdly, 
  obtain the following equation for the scale factor:
\be
  {\mathfrak I} - H\frac{\partial {\mathfrak I}}{\partial H} +
  \left(-\dot H + 3 H^2\right) \frac{\partial {\mathfrak I}}{\partial
    \dot H} + H \frac{d}{dt} \frac{\partial {\mathfrak I}}{\partial
    \dot H} = 0~, \la{genEqn} \ee
which is just the generalisation of the Friedmann equation.
\end{itemize}
\underline{\sl Remark 3}\\
Equation~\eqref{genEqn} is third order in
$a$, while the equation of motion $\delta W/\delta a =0$ is of
fourth order. One can easily check that the above prescription
is just a formulation of energy conservation.  Indeed, since
$W[a(t)]$ in Eq.~\eqref{Form} does not depend explicitly on time, 
Nother's theorem implies conservation of the quantity:
\be E \equiv\frac{\partial L}{\partial a_{t}}a_{t} - L +
\frac{\partial L}{\partial a_{tt}} a_{tt} - \left(\frac{d}{d
  t}\frac{\partial L}{\partial a_{tt}}\right) ~. \la{encons} \ee
If in addition, one imposes that the overall (gravity+fields) energy
$E$ vanishes, then the resulting equation will be exactly Eq.~\eqref{genEqn}.

In our case the action is given by\footnote{The cosmological constant $\lambda$  is known to be much smaller
than all other constants of dimension four, in particular $M_{\rm Pl}^4$. We do not expect that $\lambda$ is generated dynamically through
quantum anomalies and we can make no comment on its origin. In the following, we are interested to check that the cosmological constant will not destabilise the inflationary solution.}
\be W_{\rm total}\left[g_{\mu\nu}\right] = W + W_{\lambda},\quad W_{\lambda} \equiv \int {\rm d}^4 x \sqrt{-g} \left(-\lambda \right)
\la{WM} \ee
where\footnote{In what follows we skip the index $\mathbf M$ for
  brevity, keeping in mind, that we are working in  a Minkowski
  space-time.} $W$ is a quantum effective action $\frac{1}{i} \log Z$ with $Z$  defined by Eq.~\eqref{Ztot} and spectral regularisation. 

Following the prescription described above, we must substitute the
conformally flat anzatz for the metric tensor in comoving coordinates
in Eq.~\eqref{Scollansw}. This is done in two steps: firstly, we substitute the conformally flat anzatz in the conformal coordinates $g_{\mu\nu} =
e^{2\beta(\tau)}\eta_{\mu\nu}$ and secondly, we perform the
corresponding change of variables to the comoving frame. Hence, substituting
$g_{\mu\nu} = \eta_{\mu\nu}$ and $\phi = \beta(\tau)$ in Eq.~\eqref{Scollansw}, we get\footnote{We use the fact, that $W[\eta_{\mu\nu}] = 0$, that can be easily checked by direct computation, since in this
case the  spectrum of each Laplacian, appearing under the sign of determinant is trivial.} 
\be W_{\rm total}\left[\beta(\tau)\right] = {\rm vol}\cdot \int {\rm d}\tau\left( 3
\alpha_2 e^{2\beta}\beta_{\tau}^2 + \left(3\alpha_5 -
2\alpha_4\right)\beta_{\tau}^4 + 3\alpha_5\beta_{\tau\tau}^2 - \lambda
e^{4\beta}\right). \la{Wconf} \ee
Performing the change of variables $\beta(\tau)\rightarrow a(t)$ with
\be \tau = \int_{t_0}^t
a^{-1}\left(z\right)dz, \quad \beta(\tau) = \log{a(t)}, \ee
we arrive to the following expression for the effective action:
\be
W_{\rm total}[(a(t))] = {\rm vol} \cdot \int dt~ a^3 ~{\mathfrak I}\left(H,\dot
H\right)~,
\ee
where
\be
{\mathfrak I}(H,\dot H)
\equiv 3 \alpha_2 H^2 + \left(6 \alpha_5 - 2 \alpha_4\right)H^4 + 3
\alpha_5 \dot H^2 + 6 \alpha_5 H^2 \dot H \la{J} - \lambda~. \ee
Substituting the above  expression for $\mathfrak{J}$, Eq.~\eqref{J}, in 
Eq.~\eqref{genEqn}, we arrive to the following equation
of motion in terms of the Hubble parameter $H$: 
\be \la{eqH} 
\ddot{ H} + 3\,H\dot H
- \frac{1}{2}\,{\frac {{\dot H}^{2}}{H}} + \frac{3}{4}\,{\frac
  {{H}^{3}}{Q}} - 3\,H{ \Lambda}^{2} + {\frac {\lambda\, P}{H}} = 0~,
\ee
where
\ba
Q  &\equiv& \frac{N_{\rm F}-4N_{\rm V}}{N_{\rm F}+8 N_{\rm
    V}}~,\nonumber\\
P &\equiv& \frac{96\pi^2}{N_{\rm F} - 4N_{\rm
    V}}~. \ea
Equation \eqref{eqH} for the Hubble parameter $H(t)$ is of second order, so we write it as a system of two first order equations, in order to use the phase portrait
technique. Hence,
\be
\!\!\frac{{\rm d}}{{\rm d}t} \left(\begin{array}{c}  
           v  \\  
           H 
         \end{array}\right)  = 
         \left(\begin{array}{c} -3\,Hv+\frac{1}{2}\,{\frac
       {{v}^{2}}{H}} - \frac{3}{4}\,{\frac {{H}^{3}}{Q}}+3\,H{
       \Lambda}^{2}-{\frac {\lambda\,P}{H}} \\ v
         \end{array}\right)~.  \la{fos}
\ee
We are looking for special points of the vector field on the $[H,v]$-plane defined by the right-hand side of Eq.~\eqref{fos}. Solving this algebraic equation we find two special points\footnote{There are four special points, but since we
are interested in expanding solutions we only consider positive values of
$H$.}, $H_1$ and $H_2$:
\ba H_1&=& \sqrt{2Q}\Lambda~\sqrt{1-\sqrt{1-\frac{\lambda P}{3\Lambda^4
Q}}}\nonumber\\
&\simeq& \frac{1}{\Lambda} \sqrt {\frac{\lambda P}{3}}\left(1+O \left(
\frac{\lambda}{\Lambda^4} \right)\right)~, \la{H1} \ea
describing a slowly expanding universe, and
\ba 
H_2 &=& \sqrt{2Q}\Lambda~\sqrt{1+\sqrt{1-\frac{\lambda P}{3\Lambda^4
Q}}}\nonumber\\
&\simeq& 2\sqrt{Q}\Lambda\left(1+O \left( \frac{\lambda}{\Lambda^4}
\right)\right)~, \la{H2} 
\ea
describing a rapidly expanding universe and hence offering a good
candidate for an inflationary model.

Linearising the system Eq.~\eqref{fos} in the vicinity of each special
point, we draw the following conclusions:
\begin{itemize}
\item{The rapidly expanding solution $H_2$ is stable (stable focus on 
$[H,V]$ plane). }
\item{The slowly expanding solution $H_1$ is unstable (unstable focus
on $[H,V]$ plane).}
\end{itemize}
\begin{figure}
\includegraphics[width=0.48\linewidth]{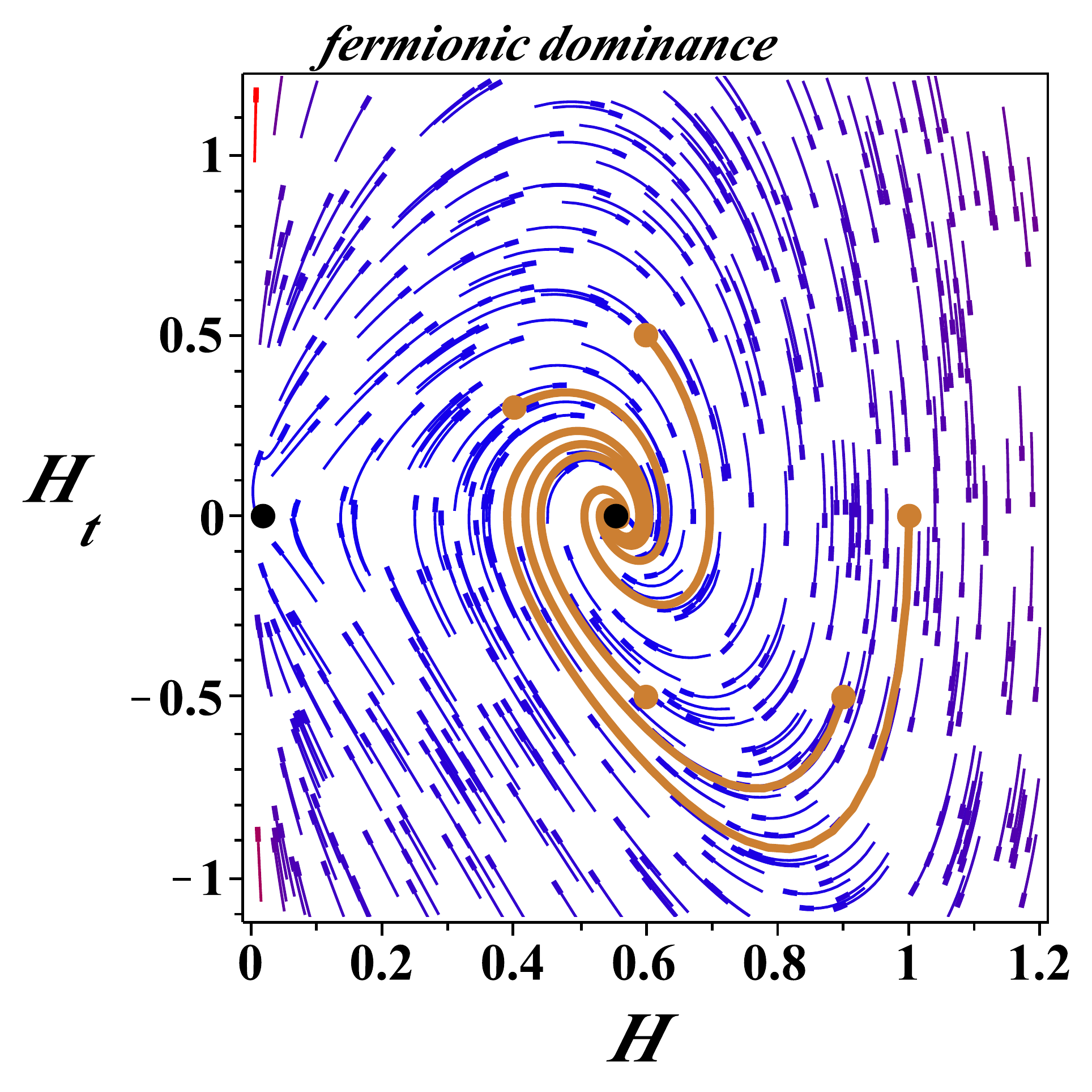}
\caption{Phase portrait showing the dynamics of the scale factor in
the case of the Sakharov's induced gravity with the Pauli compensation condition
for the quartic divergences cancellation.  The parameters are taken as
follows, $N_{\rm V} = 12$, $N_{\rm V} = 5N_{\rm V}$.This system shows the existence of a
stable de Sitter solution with the scalar curvature smaller or equal to the $M_{\rm Pl}$;
corresponds to a rapidly expanding universe. }
\end{figure}

In conclusion, if Pauli compensation condition is satisfied, namely if all quartic divergences cancel each other, then Sakharov's induced gravity leads to Starobinsky's anomaly-induced inflation, and vice versa.

\section{Conclusions \la{sec:conclusions}}
The aim of this study was to relate Sakharov's induced gravity to
the anomaly-induced effective action and thus obtain Starobinky's
anomaly-induced inflation. Imposing spectral regularisation with the cutoff
scale $\Lambda$ in a classically Weyl invariant theory, we computed the anomaly and expressed the anomalous part of the quantum effective action through the
quantised single collective scalar degree of freedom of all quantum
vacuum fluctuations, dubbed as the collective dilaton field $\phi$,
described by the local action Eq.~\eqref{Scollansw}.

It is worth noting that the condition of stability of the cosmological constant under $\Lambda^4$-corrections, namely
$N_{\mathcal H} = 2\left(N_{\rm F}^{\mathbf{w}} -N_{\rm V}\right)$, appears naturally within our procedure.

Our approach allowed us to treat the Sakharov's induced gravity on equal footing with the Starobinsky's anomaly-induced action, in a self-consistent way.  More precisely, we found that
\ba
M_{\rm Pl}^{2~\mathbf{ind}}&=&\frac{\Lambda^2}{12\pi}\left({N_{\rm F}^{\mathbf{w}}}
- 4N_{\rm V}\right)~,\no H_{\textbf{inflat}}& \simeq&
2\sqrt{\frac{N_{\rm F}^{\mathbf{w}}-4N_{\rm V}}{N_{\rm F}+8 N_{\rm V}}}\cdot\Lambda\cdot\left(1+O \left(
\frac{\lambda}{\Lambda^4} \right)\right)~. \nonumber \ea
Provided the stability condition is satisfied, Sakharov's induced
gravity and anomaly induced action leading to Starobinsky's anomaly-induced inflation appear 
simultaneously if $N_{\rm F}^{\mathbf{w}} > 4 N_{\rm V}$.

The fact that QFT gave rise to the Einstein-Hilbert
action and the onset of an inflationary era, in the absence of an inflaton field,  may indicate that the
cosmological arrow of time results from quantum effects in a classically
Weyl-invariant theory. At this point, it is worth emphasising that while our model can account for the onset of inflation and the first inflationary stage, it cannot describe the graceful exit and the subsequent evolution of the universe to its present state. In order to explain the graceful exit one should go beyond
our massless particle approximation, as was done for intance in Ref.~\cite{Shapiro:2002xd}.  
\begin{appendix}

\renewcommand{\theequation}{\Alph{section}.\arabic{equation}}
\setcounter{section}{0}
\setcounter{equation}{0}
\section{Gauge-Ghost's Contribution: Some Computational Details\label{sec:app2}}
\setcounter{equation}{0}
Starting from the Maxwell action for gauge fields  
\be
S_{\mathbf{M}} = \int {\rm d}^4 x \sqrt{g}\left(\frac{1}{4}F{_{\mu\nu}}F^{\mu\nu}\right)~,
\ee
we perform a Faddeev-Popov quantisation procedure, adding the gauge fixing term 
\be
S_{\mathbf{gf}}\equiv \frac{1}{2}\int {\rm d}^4 x \sqrt{g}\left(\nabla_{\mu}A^{\mu}\right)~,
\ee
and the ghost action $S_{\mathbf{gh}}$
\be
S_{\mathbf{gh}} = \int {\rm d}^4 x \sqrt{g} \bar c D_{\mathbf{gh}} c~,
\ee
where
\be
D_{\mathbf{gh}}   \equiv -\nabla^2~.
\ee
The overall gauge fixed Maxwell-ghost action then reads
\ba
S_{\mathbf{vec-gh}} &\equiv& S_{\mathbf{M}} + S_{\mathbf{gf}} + S_{\mathbf{gh}} \nonumber\\ 
&=& \frac{1}{2}\int {\rm d}^4 x \sqrt{g}\left(A^{\mu}(D_{\mathbf{vec}})_{\mu~}^{~\nu}A_{\nu}\right)  + \int {\rm d}^4 x \sqrt{g} \bar c D_{\mathbf{gh}} c~,
\ea
where
\bea
(D_{\mathbf{vec}})^{\nu}_{~\mu} &\equiv& - \delta_{\mu}^{\nu}\nabla^2 - \left[\nabla_{\mu},\nabla^{\nu}\right] \nonumber\\
&=& - \delta_{\mu}^{\nu}\nabla^2 - R^{\nu}_{\mu}~.
\ea
The partition function describing a contribution of the quantised vector fields and ghosts to the vacuum energy is given by the functional integral
\ba
Z_{\mathbf{vec-gh}} &=& \int [{\rm d} A][{\rm d}\bar c][{\rm d} c]e^{- S_{\mathbf{vec-gh}}}\nonumber\\
& =& \frac{\det{D_{\mathbf{gh}}}}{\sqrt{\det{D_{\mathbf{vec}}}}}~. \la{Zvg} 
\ea
Since both operators $ D_{\mathbf{vec}}$ and $D_{\mathbf{gh}}$ are unbounded, the last equality is formal and thus we perform a spectral regularisation.
Following the same approach as for the fermonic and scalar cases, we introduce the cutoff scale $\Lambda$ and the two projectors
\ba
P_{\mathbf{vec}}^{\Lambda} &=& \Theta\left(\Lambda^2 - D_{\mathbf{vec}}^2\right)~, \no
P_{\mathbf{gh}}^{\Lambda} &=& \Theta\left(\Lambda^2 - D_{\mathbf{gh}}^2\right)~,
\ea
in order to truncate the spectrum of the $ D_{\mathbf{vec}}$ and $D_{\mathbf{gh}}$ operators, respectively.

The regularisation is based on replacing the unbounded operators
$D_{\mathbf{vec}}$ and $D_{\mathbf{gh}}$ by the truncated operators
$D_{\mathbf{vec}}^{\Lambda}$ and $D_{\mathbf{gh}}^{\Lambda}$, respectively, denoted by
\ba
D_{\mathbf{vec}} &\rightarrow& D_{\mathbf{vec}}^{\Lambda}\equiv\left(\frac{D_{\mathbf{vec}}}{\Lambda}\right)P_{\mathbf{vec}}^{\Lambda} + 1-P_{\mathbf{vec}}^{\Lambda}~, \no
D_{\mathbf{gh}} &\rightarrow& D_{\mathbf{gh}}^{\Lambda}\equiv \left(\frac{D_{\mathbf{gh}}}{\Lambda}\right)P_{\mathbf{gh}}^{\Lambda} + 1-P_{\mathbf{gh}}^{\Lambda}~,
\ea
in the determinants appearing in the partition function Eq.~\eqref{Zvg}.
Hence, the regularised partition function reads
\ba
Z_{\mathbf{vec-gh}}^{\Lambda} &\equiv& \frac{\det{D^{\Lambda}_{\mathbf{gh}}}}{\sqrt{\det{D^{\mathbf{\Lambda}}_{\mathbf{vec}}}}}\nonumber\\
&=& -\exp\left\{\Tr\left({\frac{1}{2}P_{\mathbf{vec}}^{\Lambda}\log{D_{\mathbf{vec}}}}\right) - \Tr\left({P_{\mathbf{gh}}^{\Lambda}
\log{D_{\mathbf{gh}}}}\right)\right\}~. \la{ZvgReg}
\ea
We are interested in computing the contribution of the quantised vector fields and ghosts to the anomaly $S_{\mathbf{coll}}$.
We will impose the above discussed regularisation and use Eq.~\eqref{ZvgReg} for the regularised partition function.
Note that we denote a quantity $Q[g_{\mu\nu}]$ computed on the transformed metric tensor $g_{\mu\nu}e^{2\phi}$ by 
\be\left(\widetilde{Q}\right)_{\phi}\equiv Q\left[e^{2\phi}g_{\mu\nu}\right]~.\ee
For the anomaly we have
\ba\label{expansion}
S_{\mathbf{coll}} &=& \log{\left(\frac{\left(\widetilde{Z_{\mathbf{vec-gh}}^{\Lambda}}\right)_{\phi}}{Z_{\mathbf{vec-gh}}^{\Lambda}}\right)} \nonumber\\
&=& \int_0^{1} {\rm d}t ~ \partial_t \log{\left(\widetilde{Z_{\mathbf{vec-gh}}^{\Lambda}}\right)_{\phi\cdot t}}  \label{Scollcomput}\nonumber\\
&=&-\int_0^{1} {\rm d}t ~ \partial_t  
\left\{\Tr\left({\frac{1}
{2}\widetilde{P_{\mathbf{vec}}^{\Lambda}}\log{ \widetilde{D_{\mathbf{vec}}}   }}\right)_{\phi\cdot t}
 - \Tr\left({\widetilde{P_{\mathbf{gh}}^{\Lambda}}
\log{\widetilde{D_{\mathbf{gh}}}  }}\right)_{\phi\cdot t}\right\} \no
 &=& -\int_0^1 {\rm d}t \left( \frac{1}{2}
\Tr\left[\widetilde{P_{\mathbf{vec}}^{\Lambda}} 
\left(\widetilde{D_{\mathbf{vec}}^{\Lambda}}\right)^{-1} 
 \partial_t \widetilde{D_{\mathbf{vec}}^{\Lambda}}\right]_{\phi\cdot t} 
\right.
\nonumber\\
&& \ \ \ \ \left.
\ \ \ \ \ \ \ \ - \Tr\left[\widetilde{P_{\mathbf{gh}}^{\Lambda}} \left(\widetilde{D_{\mathbf{gh}}^{\Lambda}}\right)^{-1} 
 \partial_t \widetilde{D_{\mathbf{gh}}^{\Lambda}}\right]_{\phi\cdot t}\right)~.
\ea
In contrast to the fermonic and scalar cases, the next step in the computation of the anomaly $S_{\mathbf{coll}}$ is not a trivial task, because both operators $D_{\mathbf{dec}}$ and $D_{\mathbf{gh}}$ do not transform in a covariant way, namely
\ba
\left[\widetilde{\left(D_{\mathbf{vec}}\right)}_{\mu}^{~\lambda} \right]_{\phi} &=& e^{-2\phi }\left({\left(D_{\mathbf{vec}}\right)}_{\mu}^{~\lambda} 
+ 2\phi_{\mu}\nabla^{\lambda} - 2\phi^{\lambda}\nabla_{\mu} - 2\phi_{\mu}^{~\lambda} + 4\phi_{\mu}\phi^{\lambda}\right), \no
\left[\widetilde{D_{\mathbf{gh}}}\right]_{\phi} &=& e^{-2\phi }\left({D_{\mathbf{gh}}} - 2\phi^{\mu}\nabla_{\mu} \right)~. \la{DvgTrans}
\ea 
From the transformation law, Eq.~\eqref{DvgTrans} above, we derive
\ba
\partial_t \left[\widetilde{\left(D_{\mathbf{vec}}\right)}_{\mu}^{~\lambda} \right]_{\phi\cdot t} &=& -2 \phi\left[\widetilde{\left(D_{\mathbf{vec}}\right)}_{\mu}^{~\lambda} \right]_{\phi\cdot t}
+2\left[\widetilde{\phi_{\mu}\nabla^{\lambda}}\right]_{\phi\cdot t}   -2 \left[\widetilde{\phi^{\lambda}\nabla_{\mu}}\right]_{\phi\cdot t}  -2\left[\widetilde{\phi_{\mu}^{~\lambda}}\right]_{\phi\cdot t}~, \no
\partial_t \left[\widetilde{  D_{\mathbf{gh}}  } \right]_{\phi\cdot t} &=& -2 \phi\left[\widetilde{  D_{\mathbf{gh}}  } \right]_{\phi\cdot t}  -2\left[\widetilde{\phi^{\mu}\nabla_{\mu}}\right]_{\phi\cdot t}~.
\la{derivatives}
\ea
Substituting Eq.~\eqref{derivatives} in Eq.~\eqref{Scollcomput} we
obtain the expression for the anomaly; it has a part similar
to that of the fermonic and bosonic cases and in addition there are
some ``bad terms", namely \be S_{\mathbf{coll}} = \int_0^1 {\rm d}t
\left\{\Tr_{\mathbf{vec}}{\left(\phi\left[\widetilde{P_{\mathbf{vec}}^{\Lambda}}\right]_{\phi\cdot
    t}\right)} -
2\Tr_{\mathbf{gh}}{\left(\phi\left[\widetilde{P_{\mathbf{gh}}^{\Lambda}}\right]_{\phi\cdot
    t}\right)} + \widetilde{\left(\mbox{``bad
    terms"}\right)_{\phi\cdot t}} \right\}~, \ee where the ``bad
terms'' are given by
\ba \mbox{``bad terms"} &\equiv& 2\left\{\Tr_{\mathbf{vec}}
\left[P_{\mathbf{vec}}^{\Lambda}\left(D_{\mathbf{vec}}\right)^{-1}
  \left(\phi_{\mu}\nabla^{\lambda} - \phi^{\lambda}\nabla_{\mu} -
  \phi_{\mu}^{~\lambda}\right)\right]\right. \no &&\ \ \ \ +\left. 2
\Tr_{\mathbf{gh}}\left[P_{\mathbf{gh}}^{\Lambda}
  \left(D_{\mathbf{gh}}\right)^{-1}\left(\phi^{\mu}\nabla_{\mu}\right)\right]\right\}~.
\ea
In what follows we will show that the ``bad terms" cancel.
\\
Let us first introduce a complete set ${\Phi_n,~n=1,2...}$ of orthonormal eigenfunctions of the ghost operator $D_{\mathbf{gh}}$, as
\ba
D_{\mathbf{gh}}\Phi_{n} = \lambda_n\Phi_{n}\quad
\mbox{with}\ \int {\rm d}^4 x \sqrt{g} \Phi_n \Phi_m = \delta_{nm}~.
\ea 
One can easily check that the set of functions $\xi_{n}^{\mu}\equiv \frac{\nabla^{\mu}\Phi_n}{\sqrt{\lambda_n}}$ satisfies
\ba
\left(D_{\mathbf{vec}}\right)^{\mu}_{~\nu}\xi_n^{\nu} = \lambda_n \xi^{\mu}_n \quad
\mbox{with}\  \int {\rm d}^4 x \sqrt{g} \xi_{n\mu} \xi^{\mu}_m = \delta_{nm},
\ea
namely it forms an orthonormal basis in a space of longitudinal eigenvectors of the the operator $D_{\mathbf{vec}}$.
Let us also introduce the orthonormal set of transversal eigenvectors of 
\be
\left(D_{\mathbf{vec}}\right)^{\mu}_{~\nu}B_n^{\nu} = \beta_n B^{\mu}_n~ \quad
\mbox{with} \ \quad \int {\rm d}^4 x \sqrt{g} B_{n\mu} B^{\mu}_m = \delta_{nm}~ \quad \mbox{and}\  \quad \nabla_{\mu}B^{\mu}_n = 0~,
\ee
so the set $\{\xi^{\mu}_n, B^{\mu}_m\}$ with $~n,m=1, 2, \cdots$ forms a basis in the space of all gauge potentials. The gauge contribution to the ``bad terms" is
\ba
&&
\Tr
 \left[P_{\mathbf{vec}}^{\Lambda}\left(D_{\mathbf{vec}}\right)^{-1} \left(\phi_{\mu}\nabla^{\lambda} 
- \phi^{\lambda}\nabla_{\mu} - \phi_{\mu}^{~\lambda}\right)\right] \no
&=& \sum_{n:~\lambda_n \leq\Lambda}\int {\rm d}^4 x \sqrt{g}\left(\xi_n^{\mu} \left(D_{\mathbf{vec}}^{-1}\right)_{\mu}^{~\eta} \left(\phi_{\eta}\nabla^{\lambda} 
- \phi^{\lambda}\nabla_{\eta} - \phi_{\eta}^{~\lambda}\right)\xi_{n\lambda}\right)\no
&&+ \sum_{n:~\beta_n \leq\Lambda}\underbrace{\int {\rm d}^4 x \sqrt{g}\left(B_n^{\mu} \left(D_{\mathbf{vec}}^{-1}\right)_{\mu}^{~\eta} \left(\phi_{\eta}\nabla^{\lambda} 
- \phi^{\lambda}\nabla_{\eta} - \phi_{\eta}^{~\lambda}\right)B_{n\lambda}\right)}_0\no
&=& -2  \sum_{n:~\lambda_n \leq\Lambda}\frac{1}{\lambda_n}\int {\rm d}^4 x \sqrt{g} \left(\Phi_n\phi^{\nu}\nabla_{\nu}\Phi_n\right)~,
\ea
and the ghost contribution to the ``bad terms" reads
\ba
&&2
\Tr
\left[P_{\mathbf{gh}}^{\Lambda}\left(D_{\mathbf{gh}}\right)^{-1}\left(\phi^{\mu}\nabla_{\mu}\right) \right] \nonumber\\
=&&2  \sum_{n:~\lambda_n \leq\Lambda}\frac{1}{\lambda_n}\int {\rm d}^4 x \sqrt{g} \left(\Phi_n\phi^{\nu}\nabla_{\nu}\Phi_n\right)\no
=&& - 
\Tr
 \left[P_{\mathbf{vec}}^{\Lambda}\left(D_{\mathbf{vec}}\right)^{-1} \left(\phi_{\mu}\nabla^{\lambda} 
- \phi^{\lambda}\nabla_{\mu} - \phi_{\mu}^{~\lambda}\right)\right]~.
\ea
Clearly, the ``bad terms" cancel mode by mode.

Hence, the final answer for the gauge-ghost contribution to the anomaly is
\be
S_{\mathbf{coll}} =  \int_0^1 {\rm d}t 
\left\{\Tr
{\left(\phi\left[\widetilde{\chi\left(\frac{D_{\mathbf{vec}}}{\Lambda^2}\right)}\right]_{\phi\cdot t}\right)}  - 
2\Tr
{\left(\phi\left[\widetilde{\chi\left(\frac{D_{\mathbf{gh}}}{\Lambda^2}\right)}\right]_{\phi\cdot t}\right)}  \right\}~,
\ee
where the cutoff function $\chi$ is the Heaviside step-function
\be
\chi(z)\equiv \Theta(1-z)~.
\ee

\end{appendix}

\flushleft{\bf Acknoweledgments}

It is a pleasure to thank Fedele Lizzi for fruitful discussions. The work of M.\ K. was partially supported by the RFBR grant 13-02-00127-a.


\begin{thebibliography}{99}
\bibitem{Starobinsky:1980te}
  A.~A.~Starobinsky,
  Phys.\ Lett.\ B {\bf 91} (1980) 99.

\bibitem{Guth:1980zm}
  A.~H.~Guth,
  Phys.\ Rev.\ D {\bf 23} (1981) 347.
  
\bibitem{Linde:1981mu}
  A.~D.~Linde,
  Phys.\ Lett.\ B {\bf 108} (1982) 389.

\bibitem{Calzetta:1992gv}
  E.~Calzetta and M.~Sakellariadou,
  Phys.\ Rev.\ D {\bf 45},  2802 (1992).

\bibitem{Calzetta:1992bp}
  E.~Calzetta and M.~Sakellariadou,
  Phys.\ Rev.\ D {\bf 47},  3184 (1993)
  [gr-qc/9209007].

\bibitem{Germani:2007rt}
  C.~Germani, W.~Nelson and M.~Sakellariadou,
  Phys.\ Rev.\ D {\bf 76},  043529  (2007)
  [gr-qc/0701172 [gr-qc]].



\bibitem{Bezrukov:2007ep}
  F.~L.~Bezrukov and M.~Shaposhnikov,
  Phys.\ Lett.\ B {\bf 659} (2008) 703
  [arXiv:0710.3755 [hep-th]].

\bibitem{Burgess:2009ea}
  C.~P.~Burgess, H.~M.~Lee and M.~Trott,
  JHEP {\bf 0909} (2009) 103
  [arXiv:0902.4465 [hep-ph]]

\bibitem{Barbon:2009ya}
  J.~L.~F.~Barbon and J.~R.~Espinosa,
  Phys.\ Rev.\ D {\bf 79} (2009) 081302
  [arXiv:0903.0355 [hep-ph]].
 
\bibitem{Buck:2010sv}
  M.~Buck, M.~Fairbairn and M.~Sakellariadou,
  Phys.\ Rev.\ D {\bf 82} (2010) 043509
  [arXiv:1005.1188 [hep-th]].

\bibitem{Atkins:2010yg}
  M.~Atkins and X.~Calmet,
  Phys.\ Lett.\ B {\bf 697} (2011) 37
  [arXiv:1011.4179 [hep-ph]].


\bibitem{Rocher:2004et}
  J.~Rocher and M.~Sakellariadou,
  JCAP {\bf 0503}, 004 (2005) 
  [hep-ph/0406120].

\bibitem{Rocher:2004my} 
  J.~Rocher and  M.~Sakellariadou,
  Phys.\ Rev.\ Lett.\  {\bf 94}, 011303 (2005)
  [hep-ph/0412143].
 

\bibitem{Rocher:2006nh}
  J.~Rocher and M.~Sakellariadou,
  JCAP {\bf 0611},  001 (2006)
  [hep-th/0607226].


\bibitem{Battye:2010hg}
  R.~Battye, B.~Garbrecht and A.~Moss,
  Phys.\ Rev.\ D {\bf 81},  123512 (2010)
  [arXiv:1001.0769 [astro-ph.CO]].

\bibitem{Jeannerot:2003qv}
  R.~Jeannerot, J.~Rocher and M.~Sakellariadou,
  Phys.\ Rev.\ D {\bf 68}, 103514  (2003) 
  [hep-ph/0308134].

\bibitem{Ade:2013xla}
  P.~A.~R.~Ade {\it et al.}  [Planck Collaboration],
  arXiv:1303.5085 [astro-ph.CO].

\bibitem{Cacciapaglia:2013tga}
  G.~Cacciapaglia and M.~Sakellariadou,
  arXiv:1306.3242 [hep-ph].

\bibitem{Kofman}
L.\ A.\ Kofman, A.\ D.\ Linde and A.\ A. Starobinsky,
Phys.\ Lett.\ B {\bf 157} (1985) 361.

\bibitem{Gottlober:1990um}
  S.~Gottlober, V.~Muller and A.~A.~Starobinsky,
  Phys.\ Rev.\ D {\bf 43} (1991) 2510.

\bibitem{Vilenkin:1985md}
  A.~Vilenkin,
  Phys.\ Rev.\ D {\bf 32} (1985) 2511.

\bibitem{Shapiro:2003gm}
  I.~L.~Shapiro,
  Nucl.\ Phys.\ Proc.\ Suppl.\  {\bf 127} (2004) 196
  [hep-ph/0311307].

\bibitem{Pelinson:2002ef}
  A.~M.~Pelinson, I.~L.~Shapiro and F.~I.~Takakura,
  Nucl.\ Phys.\ B {\bf 648} (2003) 417
  [hep-ph/0208184].

\bibitem{Pelinson:2010yr}
  A.~M.~Pelinson and I.~L.~Shapiro,
  Phys.\ Lett.\ B {\bf 694} (2011) 467
  [arXiv:1005.1313 [hep-th]].

\bibitem{Fabris:2011qq}
  J.~C.~Fabris, A.~M.~Pelinson, F.~de O.Salles and I.~L.~Shapiro,
  JCAP {\bf 1202} (2012) 019
  [arXiv:1112.5202 [gr-qc]].

\bibitem{Sakharov:1967pk}
  A.~D.~Sakharov,
  Sov.\ Phys.\ Dokl.\  {\bf 12} (1968) 1040
   [Dokl.\ Akad.\ Nauk Ser.\ Fiz.\  {\bf 177} (1967) 70]
   [Sov.\ Phys.\ Usp.\  {\bf 34} (1991) 394]
   [Gen.\ Rel.\ Grav.\  {\bf 32} (2000) 365].

\bibitem{Visser:2002ew}
  M.~Visser,
  Mod.\ Phys.\ Lett.\ A {\bf 17} (2002) 977
  [gr-qc/0204062].

\bibitem{jeta-functional}
J.\ S.\ Dowker and R.\ Critchley,
Phys.\ Rev.\ D {\bf 13} (1976) 3224.

\bibitem{Schwinger}
J.\ Schwinger, Phys.\ Rev.\ D {\bf 82} (1951) 664.

\bibitem{Riegert:1984kt}
  R.~J.~Riegert,
  Phys.\ Lett.\ B {\bf 134} (1984) 56.

\bibitem{Andrianov:1983fg}
  A.~A.~Andrianov and L.~Bonora,
  Nucl.\ Phys.\ B {\bf 233} (1984) 232.

\bibitem{Andrianov:1987jh}
  A.~A.~Andrianov, V.~A.~Andrianov, V.~Y.~Novozhilov and Y.~V.~Novozhilov,
  Phys.\ Lett.\ B {\bf 186} (1987) 401.

\bibitem{Vassilevich:1987yn}
  D.~V.~Vassilevich and Y.~.V.~Novozhilov,
  Theor.\ Math.\ Phys.\  {\bf 73} (1987) 1237
   [Teor.\ Mat.\ Fiz.\  {\bf 73} (1987) 308].
 
\bibitem{Andrianov:2011bc}
  A.~A.~Andrianov, M.~A.~Kurkov and F.~Lizzi,
  JHEP {\bf 1110} (2011) 001
  [arXiv:1106.3263 [hep-th]].

\bibitem{Kurkov:2012dn}
  M.~A.~Kurkov and F.~Lizzi,
  Mod.\ Phys.\ Lett.\ A {\bf 27} (2012) 1250203
  [arXiv:1210.2663 [hep-th]].

\bibitem{Andrianov:1985ay} 
  A.~A.~Andrianov,
  Phys.\ Lett.\ B {\bf 157} (1985) 425.

\bibitem{Vassilevich:2003xt}
  D.~V.~Vassilevich,
  Phys.\ Rept.\  {\bf 388} (2003) 279
  [hep-th/0306138]. 

\bibitem{Birrell:1982ix}
  N.~D.~Birrell and P.~C.~W.~Davies,
  ``Quantum Fields in Curved Space,''
  Cambridge Monogr.Math.Phys. Cambridge University Press 1982.
 
\bibitem{Andrianov:2010nr}
  A.~A.~Andrianov and F.~Lizzi,
  JHEP {\bf 1005} (2010) 057
  [arXiv:1001.2036 [hep-th]].

\bibitem{Chamseddine:1996zu}
  A.~H.~Chamseddine and A.~Connes,
  Commun.\ Math.\ Phys.\  {\bf 186} (1997) 731
  [hep-th/9606001].
 
\bibitem{ncg-book1} A.\ Connes, {\sl Noncommutative Geometry},
  Academic Press, New York (1994).

\bibitem{ncg-book2}  A.\ Connes and M.~Marcolli, {\sl Noncommutative Geometry,
 Quantum Fields and Motives}, Hindustan Book Agency, India (2008).

\bibitem{ccm} A.~H.~Chamseddine, A.~Connes and M.~Marcolli,
  Adv.\ Theor.\ Math.\ Phys.\  {\bf 11}, 991 (2007)
  [arXiv:hep-th/0610241].

\bibitem{cchiggs}
A.~H.~Chamseddine and A.~Connes,
JHEP {\bf 1209}, 104 (2012)
[arXiv:1208.1030 [hep-ph]].

\bibitem{Chamseddine:2013rta}
  A.~H.~Chamseddine, A.~Connes and W.~D.~van Suijlekom,
  [arXiv:1304.8050 [hep-th]].
  
\bibitem{Devastato:2013oqa}
  A.~Devastato, F.~Lizzi and P.~Martinetti,
  [arXiv:1304.0415 [hep-th]].

\bibitem{Nelson:2008uy}
  W.~Nelson and M.~Sakellariadou,
  Phys.\ Rev.\ D {\bf 81}, 085038 (2010)
  [arXiv:0812.1657 [hep-th]].
 
\bibitem{Nelson:2009wr}
  W.~Nelson and M.~Sakellariadou,
  Phys.\ Lett.\  B {\bf 680}, 263 (2009)
  [arXiv:0903.1520 [hep-th]].
  
\bibitem{Marcolli:2009in}
  M.~Marcolli and E.~Pierpaoli,
  [arXiv:0908.3683 [hep-th]].

\bibitem{Nelson:2010ru}
  W.~Nelson, J.~Ochoa and M.~Sakellariadou,
  Phys.\ Rev.\ Lett.\  {\bf 105} (2010) 101602
  [arXiv:1005.4279 [hep-th]].
  
\bibitem{Nelson:2010rt}
  W.~Nelson, J.~Ochoa and M.~Sakellariadou,
  Phys.\ Rev.\  D {\bf 82} (2010) 085021
  [arXiv:1005.4276 [hep-th]].

\bibitem{Lambiase:2013dai}
  G.~Lambiase, M.~Sakellariadou and A.~Stabile,
  arXiv:1302.2336 [gr-qc].

\bibitem{Weinberg:2009wa}
  S.~Weinberg,
  Phys.\ Rev.\ D {\bf 81} (2010) 083535
  [arXiv:0911.3165 [hep-th]].
  
 \bibitem{Shapiro:2002xd}
  I.~L.~Shapiro and J.~Sola,
  Russ.\ Phys.\ J.\  {\bf 45} (2002) 727
   [Izv.\ Vuz.\ Fiz.\  {\bf 2002N7} (2002) 75].
  

\end{thebibliography}
\end{document}